\documentclass[runningheads]{llncs}
\usepackage[T1]{fontenc}
\usepackage{graphicx}
\usepackage[table]{xcolor}
\usepackage[english]{babel}
\usepackage[utf8]{inputenc}
\usepackage{eurosym}
\usepackage{float}
\usepackage[T1]{fontenc}
\usepackage{geometry}
\usepackage{bm}
\usepackage{amsmath}
\usepackage{babel}
\begin{document}
\title{A Geometric Procedure for Computing Differential Characteristics of Multi-phase Electrical Signals using Geometric Algebra}

%%%%%
\author{Ahmad H. Eid\inst{1}\orcidID{0000-0002-1999-651X} \and
Francisco G. Montoya\inst{2}\orcidID{0000-0003-4105-565X}}

\authorrunning{A.H. Eid and F.G. Montoya}

\institute{Port Said University, Port Said Egypt 
\\
\email{ahmad.eid@eng.psu.edu.eg}
\and
University of Almeria, 04120 Almeria, Spain\\
\email{pagilm@ual.es}}
\maketitle              % typeset the header of the contribution
\begin{abstract}
This paper presents exploratory investigations on the concept of generalized geometrical frequency in electrical systems with an arbitrary number of phases by using Geometric Algebra and Differential Geometry. By using the concept of Darboux bivector it is possible to find a bivector that encodes the invariant geometrical properties of a spatial curve named electrical curve. It is shown how the traditional concept of instantaneous frequency in power networks can be intimately linked to the Darboux bivector. Several examples are used to illustrate the findings of this work.

\end{abstract}
\section{Introduction}

Multi-phase power systems play a crucial role in modern society due to the tremendous increase in energy needs. Moreover, with the proliferation of new generation smart grids based on a decentralized paradigm and focused on renewable energies, it is of utmost importance to investigate new methodologies that can deal with the distortion and unbalance scenarios produced by nonlinear loads. In this regard, voltage and frequency control is essential to achieve adequate stability, so appropriate tools are necessary for a better understanding of transient phenomena that can potentially disturb the grid. 
Currently, the concept of instantaneous frequency, widely used in electrical power systems, presents some issues in its classical definition as the time derivative of the phase angle of a signal. Apparently, this definition only holds for a sine representation of such a signal but fails for other representations where harmonics or transients are included \cite{kirkham2018defining}. Existing techniques based on the Fourier or Hilbert Transform, for example, are unable to deal with these problems and give rise to a number of paradoxes, described in \cite{leon1995time}.

Geometric Algebra (GA) and Differential Geometry (DG) are among the most promising tools recently proposed, as evidenced by recent works \cite{montoya2021determination,montoya2022geometric}. For example, in \cite{milano2021geometrical,milano2021applications} a geometrical interpretation of the frequency for three-phase electric circuits is proposed. Through the analysis of the invariants of spatial curves, it is possible to find a direct relation with the generalized concept of frequency that can be of interest in electrical power systems. However, a generalization to an arbitrary number of phases has yet to be presented. In particular, the study of multiphase electrical machines or electrical power systems with a number of phases greater than three can benefit from the investigations presented in this paper. 

This approach not only allows a generalization through the use of exterior algebra or geometric algebra but also provides a unifying mathematical framework.

In this paper, GA and DG is applied to multi-phase electrical systems comprising any number of electrical phases by characterizing the voltage as a vector describing a curve in $n$-dimensional space. We will refer to such curves as ``electrical curves'' (EC). A new procedure is proposed that allows to obtain a multivector representation of the geometrical angular frequency, known as Darboux Bivector.

\section{Electrical Curves and Geometric Properties}

The electric curve approach is an effort for the application of concepts related to spatial curves within electrical systems. More specifically, geometric invariants (e.g. curvature) of the curve can accurately describe properties of interest for the power community. For example, the instantaneous or average grid frequency in power systems can be linked easily to curvature properties. This approach lead us to the concept of ``geometric frequency'' as introduced by Milano \cite{milano2021geometrical}. For higher dimensional, i.e., multi-phase power systems, the procedure depends on the derivations of Hestenes \cite{D.Hestenes1987} (chapter 6) where arc-length parameterization is assumed. Here we provide a procedure that expresses this idea using time-dependent formulation of the original arc-length parameterization formulation.

\subsection{Electrical Curve Definition and Parameterization}

In practical power systems applications, we are typically given a uniformly sampled multi-phase and periodic voltage (or current) signals ${v}_i\left[k\right]$ with $k$ the sample index and $i=1,2,\ldots,n$, the electrical phase index, and $n$  the total number of electrical phases. We are allowed to create a discrete vector signal $\bm{v}[k]$ from $v_i[k]$ that describes a curve in an $n$-dimensional Euclidean space. We call this object an ``electrical curve'' (EC).
 
Because the method depends on differential geometric characteristics of curves, the first step in this procedure is to use a suitable interpolation or fitting method to obtain a time-dependent differentiable curve $\boldsymbol{v}(t)$
that closely approximates the sampled signal $\boldsymbol{v}\left[k\right]$. However, real-world signals can typically contain a fair amount of noise and artifacts because of the Analog to Digital Converter (ADC) quantization sampling process, transient phenomena, etc., that make this a hard task. Fortunately, many  procedures exist in practice \cite{RichardKhoury2016,buzzi2010interpolation}. We will assume this step is already implemented using any suitable method.

Assume now a time-dependent vector $\boldsymbol{v}\left(t\right)=\sum_{i=1}^{n}v_{i}\left(t\right)\boldsymbol{\sigma}_{i}$ that describes a curve in $n$-dimensional Euclidean space defined on the interval $t\in\left[t_{0},t_{1}\right]$.
Theoretically, we can express the curve by a reparameterization $\boldsymbol{v}\left(s\right)=\boldsymbol{v}\left(t\left(s\right)\right)=\sum_{i=1}^{n}v_{i}\left(t\left(s\right)\right)\boldsymbol{\sigma}_{i}$
using the arc-length variable $s\left(t\right)=\int_{t_{0}}^{t}\left\Vert \boldsymbol{v}^{\prime}\left(\alpha\right)\right\Vert d\alpha$,
where the relation $s\left(t\right)$ and its functional inverse $t\left(s\right)=s^{-1}\left(t\right)$
between parameters $s$ and $t$ are one-to-one. Note that this reparameterization $s\left(t\right),v_{i}\left(t\left(s\right)\right)$
is not always possible to express in closed form in most cases. Therefore, in-depth new knowledge is required to overcome the exposed challenges and issues.

\subsection{Time and Arc-Length Derivatives of Electrical Curves}

%In the following derivations, we will express all computations in terms of $t$-derivatives of the given curve only, so we can avoid having to compute explicit expressions for quantities depending on $s$-derivatives. 
The computation of derivatives for the EC is of paramount importance. They can be obtained in two different ways: with respect to parameter $t$ or with respect to arc-length $s$. The relationship between  $s$ and $t$ is crucial in this regard.
 
To investigate this point, we can start by calculating the $t$-derivatives of the arc-length parameter $s\left(t\right)$ using the vector $\bm{v}$ and basic rules of vector differentiation \cite{Hubbard1999}. The following illustrates the first four derivatives:

\begin{equation}
	\begin{aligned}
		s^{\prime}\left(t\right) & =  \sqrt{\boldsymbol{v}^{\prime}\cdot\boldsymbol{v}^{\prime}}\\
		s^{\prime\prime}\left(t\right) & = \frac{1}{s^{\prime}}\boldsymbol{v}^{\prime}\cdot\boldsymbol{v}^{\prime\prime}\\
		s^{\prime\prime\prime}\left(t\right) & =  \frac{1}{s^{\prime}}\left(\boldsymbol{v}^{\prime\prime}\cdot\boldsymbol{v}^{\prime\prime}+\boldsymbol{v}^{\prime}\cdot\boldsymbol{v}^{\prime\prime\prime}-\left(s^{\prime\prime}\right)^{2}\right)\\
		s^{\prime\prime\prime\prime}\left(t\right) & =  \frac{1}{s^{\prime}}\left(3\boldsymbol{v}^{\prime\prime}\cdot\boldsymbol{v}^{\prime\prime\prime}+\boldsymbol{v}^{\prime}\cdot\boldsymbol{v}^{\prime\prime\prime\prime}-3s^{\prime\prime}s^{\prime\prime\prime}\right)\label{eq:sDt}
	\end{aligned}
\end{equation}

Having a suitable closed form time-dependent curve $\boldsymbol{v}\left(t\right)=\sum_{i=1}^{n}v_{i}\left(t\right)\boldsymbol{\sigma}_{i}$
of class $C^{p}$ (i.e. differentiable up to $p$ times in $t$),
it is simple to find the $t$-derivatives of arbitrary degree $0<m\leq p$
using:

\begin{equation}
	\begin{aligned}
	\boldsymbol{v}^{\prime}\left(t\right)=\partial_{t}\boldsymbol{v}\left(t\right) & =  \sum_{i=1}^{n}v_{i}^{\prime}\left(t\right)\boldsymbol{\sigma}_{i}\\
	\boldsymbol{v}^{\prime\prime}\left(t\right)=\partial_{t}^{2}\boldsymbol{v}\left(t\right) & =  \sum_{i=1}^{n}v_{i}^{\prime\prime}\left(t\right)\boldsymbol{\sigma}_{i}\\
	& \vdots\\
	\boldsymbol{v}^{({m})}\left(t\right)=\partial_{t}^{m}\boldsymbol{v}\left(t\right) & =  \sum_{i=1}^{n}\left[\partial_{t}^{m}v_{i}\left(t\right)\right]\boldsymbol{\sigma}_{i}
	\end{aligned}
\end{equation}
\setcounter{footnote}{0} 
On the other hand, the derivatives of the curve with respect to arc-length parameter\footnote{We use a dot for $s$-derivatives instead of a prime used in $t$-derivatives.} $s$ can be obtained in a similar fashion:

\begin{equation}
	\begin{aligned}
	\dot{\boldsymbol{v}}\left(s\right)=\partial_{s}\boldsymbol{v}\left(s\right) & =  \sum_{i=1}^{n}\dot{v}_{i}\left(s\right)\boldsymbol{\sigma}_{i}\\
	\ddot{\boldsymbol{v}}\left(s\right)=\partial_{s}^{2}\boldsymbol{v}\left(s\right) & =  \sum_{i=1}^{n}\ddot{v}_{i}\left(s\right)\boldsymbol{\sigma}_{i}\\
	\dddot{\boldsymbol{v}}\left(s\right)=\partial_{s}^{2}\boldsymbol{v}\left(s\right) & =  \sum_{i=1}^{n}\dddot{v}_{i}\left(s\right)\boldsymbol{\sigma}_{i}\\
		& \vdots\\
	\boldsymbol{v}^{({m})}\left(s\right)=\partial_{s}^{m}\boldsymbol{v}\left(s\right) & =  \sum_{i=1}^{n}\left[\partial_{s}^{m}v_{i}\left(s\right)\right]\boldsymbol{\sigma}_{i}
	\end{aligned}
\label{eq:equations_s}
\end{equation}

Interestingly, if one wants to express the above set of equations \eqref{eq:equations_s} in terms of the parameter $t$, it is found that they are much more intricate. For example, using the chain rule, the first $s$-derivative in terms of $t$ can be obtained for every $v_i(s)$:

\begin{equation}
	\dot{v}_{i}\left(s(t)\right)=\partial_{s}v_{i}\left(s(t)\right) =  \frac{dv_{i}(s(t))}{dt}\frac{dt}{ds}=\frac{1}{s^{\prime}\left(t\right)}v_{i}^{\prime}\left(t\right)
	\label{eq:chain_rule}
\end{equation}

From now on, we remove the $t$ symbol to indicate dependency on time in $s$ and $v$ for simplicity. The second and third derivatives follow:

\begin{equation}
	\begin{aligned}
		\ddot{v}_{i}\left(s\right)=\partial_{s}^{2}v_{i}\left(s\right)=\partial_{s}\dot{v}_{i}\left(s\right) & = \frac{1}{s^{\prime}}\frac{d}{dt}\left(\frac{v_{i}^{\prime}}{s^{\prime}}\right)
		 =  \frac{1}{\left(s^{\prime}\right)^{3}}\left(s^{\prime}v_{i}^{\prime\prime}-s^{\prime\prime}v_{i}^{\prime}\right)\\
		\dddot{v}_{i}\left(s\right)=\partial_{s}^{3}v_{i}\left(s\right)=\partial_{s}\left[\partial_{s}^{2}v_{i}\left(s\right)\right] & =  \frac{1}{s^{\prime}}\frac{d}{dt}\left(\frac{1}{s^{\prime}}\frac{d}{dt}\left(\frac{1}{s^{\prime}}\frac{dv_{i}}{dt}\right)\right)\\
		& =  \frac{1}{\left(s^{\prime}\right)^{5}}\left(\left[3\left(s^{\prime\prime}\right)^{2}-s^{\prime}s^{\prime\prime\prime}\right]v_{i}^{\prime}-3s^{\prime}s^{\prime\prime}v_{i}^{\prime\prime}+\left(s^{\prime}\right)^{2}v_{i}^{\prime\prime\prime}\right)
	\end{aligned}
\end{equation}

The relation between $t$-derivatives and $s$-derivatives of degree $k\geq2$ in ${v_i}$ is algebraically complicated, not at all as simple as the first derivative where $\dot{\boldsymbol{v}}\left(s\right)=\frac{1}{s^{\prime}}\boldsymbol{v}^{\prime}$.
To illustrate this complex relation between $t$-derivatives and $s$-derivatives,
the first 4 $s$-derivatives $\dot{\boldsymbol{v}}\left(s\right),\ddot{\boldsymbol{v}}\left(s\right),\dddot{\boldsymbol{v}}\left(s\right),\ddddot{\boldsymbol{v}}\left(s\right)$
in terms of $t$-derivatives are presented:

\begin{equation}
	\begin{aligned}
	\dot{\boldsymbol{v}}\left(s\right) & =  \frac{1}{s^{\prime}}\boldsymbol{v}^{\prime}\\
	\ddot{\boldsymbol{v}}\left(s\right) & =  \frac{1}{\left(s^{\prime}\right)^{3}}\left[s^{\prime}\boldsymbol{v}^{\prime\prime}-s^{\prime\prime}\boldsymbol{v}^{\prime}\right]\\
	\dddot{\boldsymbol{v}}\left(s\right) & =  \frac{1}{\left(s^{\prime}\right)^{5}}\left[\left(s^{\prime}\right)^{2}\boldsymbol{v}^{\prime\prime\prime}-3s^{\prime}s^{\prime\prime}\boldsymbol{v}^{\prime\prime}-\left(s^{\prime}s^{\prime\prime\prime}-3\left(s^{\prime\prime}\right)^{2}\right)\boldsymbol{v}^{\prime}\right]\\
	\ddddot{\boldsymbol{v}}\left(s\right) & =  \frac{1}{\left(s^{\prime}\right)^{7}}\left[\left(s^{\prime}\right)^{3}\boldsymbol{v}^{\prime\prime\prime\prime}-6\left(s^{\prime}\right)^{2}s^{\prime\prime}\boldsymbol{v}^{\prime\prime\prime}-\left(4\left(s^{\prime}\right)^{2}s^{\prime\prime\prime}-15s^{\prime}\left(s^{\prime\prime}\right)^{2}\right)\boldsymbol{v}^{\prime\prime}\right.\\
	&\left.+\left(10s^{\prime}s^{\prime\prime}s^{\prime\prime\prime}-15\left(s^{\prime\prime}\right)^{3}-\left(s^{\prime}\right)^{2}s^{\prime\prime\prime\prime}\right)\boldsymbol{v}^{\prime}\right]
	\end{aligned}
\end{equation}

Additionally, the following relations hold:

\begin{eqnarray}
	\left\Vert \dot{\boldsymbol{v}}\right\Vert  & = & 1\\
	\left\Vert \ddot{\boldsymbol{v}}\right\Vert  & = & \frac{1}{\left(s^{\prime}\right)^{2}}\sqrt{\left\Vert \boldsymbol{v}^{\prime\prime}\right\Vert ^{2}-2\frac{s^{\prime\prime}}{s^{\prime}}\left(\boldsymbol{v}^{\prime}\cdot\boldsymbol{v}^{\prime\prime}\right)+\left(s^{\prime\prime}\right)^{2}}\\
	\dot{\boldsymbol{v}}\cdot\ddot{\boldsymbol{v}} & = & 0\label{eq:vDs12_1}
\end{eqnarray}

The quantity $\left\Vert \ddot{\boldsymbol{v}}\right\Vert $ is important
as it's the base for computing the first curvature coefficient $\kappa_{1}$
of the curve $\boldsymbol{v}$ which has important implications for the geometrical frequency as illustrated later on.

\subsection{Local Orthogonal Frames in Electrical Curves}

The next step in the proposed procedure involves the application of an orthogonalization
method, such as the Gram-Schmidt process, to the $s$-derivative
vectors $\dot{\boldsymbol{v}},\ddot{\boldsymbol{v}}$ and so on. For
a symbolic expression of the $s$-derivative vectors, this is simple
to compute using either the Classical Gram-Schmidt (CGS) \cite{Bjoerck1994},
or the Geometric Algebra-based Gram-Schmidt (GAGS) \cite{hitzer2003gram}.
For practical numerical computations, however, care must be taken
when applying the CGS\textbackslash GAGS as they are highly unstable
numerically. A much more numerically stable alternative is the Modified
Gram-Schmidt (MGS) procedure \cite{Bjoerck1994,Beilina2017}, which
we found giving much better results when orthogonalizing higher (i.e.
degree 3 or higher) $s$-derivatives. In any case, we essentially compute
at each instant of time a local orthogonal frame $\left\{ \boldsymbol{u}_{1},\boldsymbol{u}_{2},\ldots,\boldsymbol{u}_{m}\right\} ,m\leq p$
from the set of $p$ local $s$-derivative vectors $\left\{ \dot{\boldsymbol{v}},\ddot{\boldsymbol{v}},\ldots,\partial_{s}^{p}\boldsymbol{v}\right\} $.
We can also normalize the frame $\left\{ \boldsymbol{u}_{1},\boldsymbol{u}_{2},\ldots,\boldsymbol{u}_{m}\right\} $
to get a fully orthonormal local frame $\left\{ \boldsymbol{e}_{1},\boldsymbol{e}_{2},\ldots,\boldsymbol{e}_{m}\right\} $
where $\boldsymbol{e}_{m}=\frac{1}{\left\Vert \boldsymbol{u}_{m}\right\Vert }\boldsymbol{u}_{m}$.

According to the chain rule presented in \eqref{eq:chain_rule},  the arc-length derivatives of the arc-length frame can be readily  computed:

\begin{eqnarray}
	\dot{\boldsymbol{e}}_{i}\left(s\right) & = & \frac{1}{s^{\prime}}\bm{e}_i^{\prime}
\end{eqnarray}

It is also interesting to find the explicit relation between vectors
$\boldsymbol{e}_{1},\boldsymbol{e}_{2}$ and vectors $\dot{\boldsymbol{v}},\ddot{\boldsymbol{v}},\boldsymbol{v}^{\prime},\boldsymbol{v}^{\prime\prime}$.
This will be useful later for expressing the ``grid angular velocity'' blade,
on which the concept of geometric frequency is based. Applying the
GAGS process to $\dot{\boldsymbol{v}},\ddot{\boldsymbol{v}}$, we
can write:

\begin{eqnarray}
	\boldsymbol{u}_{1} & = & \dot{\boldsymbol{v}}=\frac{1}{s^{\prime}}\boldsymbol{v}^{\prime}\nonumber \\
	\boldsymbol{e}_{1} & = & \frac{{\boldsymbol{u}_1}}{\left\Vert {\boldsymbol{u}_1}\right\Vert }=\dot{\boldsymbol{v}}=\frac{\boldsymbol{v}^{\prime}}{s^{\prime}} \qquad \Longrightarrow \qquad \boldsymbol{v}^{\prime} =  s^{\prime}\boldsymbol{e}_{1}\label{eq:vDt1_eq1}
\end{eqnarray}

And for the second vector we have:

\begin{equation}
	\begin{aligned}
		\boldsymbol{u}_{2} & = \ddot{\bm{v}} = \frac{1}{\left(s^{\prime}\right)^{2}}\left(\boldsymbol{v}^{\prime\prime}-s^{\prime\prime}\boldsymbol{e}_{1}\right) \qquad 
		&\Longrightarrow & \qquad \left\Vert \boldsymbol{u}_{2}\right\Vert   = u_2= \frac{1}{\left(s^{\prime}\right)^{2}}\sqrt{ ({v}^{\prime\prime})^2 -2\frac{s^{\prime\prime}}{s^{\prime}}\left(\boldsymbol{v}^{\prime}\cdot\boldsymbol{v}^{\prime\prime}\right)+\left(s^{\prime\prime}\right)^{2}} \\
		\boldsymbol{e}_{2}  &=  \frac{\boldsymbol{u}_{2}}{\left\Vert \boldsymbol{u}_{2}\right\Vert }
		=  \frac{\boldsymbol{v}^{\prime\prime}-s^{\prime\prime}\boldsymbol{e}_{1}}{\left(s^{\prime}\right)^{2}\left\Vert \ddot{\boldsymbol{v}}\right\Vert } \qquad
		&\Longrightarrow & \qquad \boldsymbol{v}^{\prime\prime}  =  s^{\prime\prime}\boldsymbol{e}_{1}+\left(s^{\prime}\right)^{2}\left\Vert \ddot{\boldsymbol{v}}\right\Vert \boldsymbol{e}_{2} 
		 =  s^{\prime\prime}\boldsymbol{e}_{1}+\sqrt{({v}^{\prime\prime}) ^{2}-2\frac{s^{\prime\prime}}{s^{\prime}}\left(\boldsymbol{v}^{\prime}\cdot\boldsymbol{v}^{\prime\prime}\right)+\left(s^{\prime\prime}\right)^{2}}\boldsymbol{e}_{2}
		 \label{eq:vDt2_1}
	\end{aligned}
\end{equation}

\subsection{Frénet-Serret Curvature Coefficients of an Electrical Curve}

The Frénet-Serret equations were formulated for three dimensions by Jean Frédéric Frénet and Joseph Alfred Serret and generalized to higher dimensions by Camile Jordan in the XIX century. They describe some dynamic properties of moving objects along curves in space by establishing a relationship between  an orthonormal frame and its derivatives that also moves along the curve. The coefficients of these equations are known as curvature coefficients $\kappa_{i},i=1,2,\ldots,n-1$ with $n$ the number of dimensions. %along with the orthonormal arc-length frame $\boldsymbol{e}_{i}$. 
They satisfy the Frénet equations \cite{D.Hestenes1987}:

\begin{equation}
	\begin{aligned}
		\dot{\boldsymbol{e}}_{1} & =  \kappa_{1}\boldsymbol{e}_{2}\\
		\dot{\boldsymbol{e}}_{2} & = -\kappa_{1}\boldsymbol{e}_{1}+\kappa_{2}\boldsymbol{e}_{3}\\
		\dot{\boldsymbol{e}}_{i} & = -\kappa_{i-1}\boldsymbol{e}_{i-1}+\kappa_{i}\boldsymbol{e}_{i+1}\\
		\dot{\boldsymbol{e}}_{n} & =  -\kappa_{n-1}\boldsymbol{e}_{n-1}
	\end{aligned}
\end{equation}

We can re-write these equations as:

\begin{equation}
	\begin{aligned}
		\dot{\boldsymbol{e}}_{1} & =  \kappa_{1}\boldsymbol{e}_{2}\\
		\dot{\boldsymbol{e}}_{2}+\kappa_{1}\boldsymbol{e}_{1} & =  \kappa_{2}\boldsymbol{e}_{3}\\
		\dot{\boldsymbol{e}}_{i}+\kappa_{i-1}\boldsymbol{e}_{i-1} & =  \kappa_{i}\boldsymbol{e}_{i+1}\\
		\dot{\boldsymbol{e}}_{n} & =  -\kappa_{n-1}\boldsymbol{e}_{n-1}
	\end{aligned}
\end{equation}

This directly leads to the solutions:

\begin{equation}
	\begin{aligned}
	\kappa_{1} & =  \dot{\boldsymbol{e}}_{1}\cdot\boldsymbol{e}_{2} \\
	\kappa_{i} & =  \left(\dot{\boldsymbol{e}}_{i}+\kappa_{i-1}\boldsymbol{e}_{i-1}\right)\cdot\boldsymbol{e}_{i+1}= \dot{\boldsymbol{e}}_{i}\cdot\boldsymbol{e}_{i+1}\\
	\end{aligned}
\end{equation}

Another simpler alternative for computing $\kappa_{i}$ is to use
the relation from \cite{Gluck1966}:

\begin{eqnarray}
	\kappa_{i}& = & \frac{\left\Vert \boldsymbol{u}_{i+1}\right\Vert }{\left\Vert \boldsymbol{u}_{i}\right\Vert }
\end{eqnarray}

Here, the vectors $\bm{u}_i$ are defined as
 \begin{equation}
 	 \boldsymbol{u}_{i}=\partial_{s}^{i}\boldsymbol{v}-\sum_{j=1}^{i-1}\frac{\left(\partial_{s}^{i}\boldsymbol{v}\right)\cdot\boldsymbol{u}_{j}}{\boldsymbol{u}_{j}\cdot\boldsymbol{u}_{j}}\boldsymbol{u}_{j}
 \end{equation}

\noindent and are computed using the CGS or MGS process, not the GAGS as before.
However,  in this work, a slightly different  expression will be used for practical applications in power systems 
where the frequency is ultimately dependent on the time variable $t$ (instead of the arc-length variable $s$) through the voltage $v(t)$. A scaled version of $\kappa_i$ will be used known as ``scaled curvature coefficient'', $k_i$:

\begin{eqnarray}
	k_{i}\left(t\right) & = & s^{\prime}\frac{\left\Vert \boldsymbol{u}_{i+1}\right\Vert }{\left\Vert \boldsymbol{u}_{i}\right\Vert }=s^{\prime}\kappa_i
\end{eqnarray}

These scaled curvature coefficients depend explicitly on the time variable $t$.

\section{The Darboux Blades}

The original Darboux bivector $\boldsymbol{\Omega}_{H}$ is described in \cite{D.Hestenes1987}. Note that we use the subscript $H$ to highlight the Hestenes definition. It contains a summary of the local differential geometric information of the electrical curve. In this work, a slight modification of this original definition is used, where we flip the order of multiplication. The rationale behind this decision is to fit the practical definitions of frequency in power systems and the recently proposed geometric frequency for particular cases, such as sinusoidal and balanced conditions.	We can use either of the following relations as the definition for the new Darboux bivector $\boldsymbol{\Omega}$:
	
	\begin{eqnarray}
		\boldsymbol{\Omega} & = & \frac{1}{2}s'\sum_{i=1}^{n}\boldsymbol{e}_{i}\wedge\dot{\boldsymbol{e}}_{i}\label{eq:db_1}\\
		& = & \sum_{i=1}^{n-1}k_{i}\boldsymbol{e}_{i}\wedge\boldsymbol{e}_{i+1}\\
		& = & s^{\prime}\sum_{i=1}^{n-1}\frac{\boldsymbol{u}_{i}\wedge\boldsymbol{u}_{i+1}}{\left\Vert \boldsymbol{u}_{i}\right\Vert ^{2}}\\
		& = & s^{\prime}\sum_{i=1}^{n-1}\boldsymbol{u}_{i}^{-1}\wedge\boldsymbol{u}_{i+1}
	\end{eqnarray}
	
	For the special case of $s'=1$, then $\bm{\Omega}=-\bm{\Omega}_H$. Using this definition for the Darboux bivector, the following important
	relation holds true: 
	
\begin{eqnarray}
	\dot{\boldsymbol{e}}_{i} & = & \boldsymbol{e}_{i}\rfloor\boldsymbol{\Omega}=-\boldsymbol{\Omega}\lfloor\boldsymbol{e}_{i}\label{eq:db_ei_eiDs}
\end{eqnarray}

For practical usage we found that it is advisable to separate the Darboux
bivector into several 2-blades $\boldsymbol{\Omega}_{i}$, which we
will call the Darboux Blades (DBs), as follows:

\begin{eqnarray*}
	\boldsymbol{\Omega} & = & \frac{1}{2}\sum_{i=1}^{n-1}\boldsymbol{\Omega}_{i}\\
	\boldsymbol{\Omega}_{1} & = & k_{1}\boldsymbol{e}_{1}\wedge\boldsymbol{e}_{2} \qquad \rightarrow \qquad	\left\Vert \boldsymbol{\Omega}_{1}\right\Vert   =  \left|k_{1}\right|\\[2ex]
	\boldsymbol{\Omega}_{i} & = & k_{i-1}\boldsymbol{e}_{i-1}\wedge\boldsymbol{e}_{i} + k_{i}\boldsymbol{e}_{i}\wedge\boldsymbol{e}_{i+1}= s^{\prime}\boldsymbol{u}_{i}^{-1}\wedge\boldsymbol{u}_{i+1} + s^{\prime}\boldsymbol{u}_{i-1}^{-1}\wedge\boldsymbol{u}_{i}\\
\end{eqnarray*}

Note that $\boldsymbol{\Omega}_{i}$ are always 2-blades (i.e. represent
planes in $n$-dimensions), while the Darboux bivector $\boldsymbol{\Omega}$
is generally a bivector, not a 2-blade (except in 3-dimensions where
all bivectors are 2-blades). 

The first DB $\boldsymbol{\Omega}_{1}$, called the grid angular velocity
blade , has special relevance for our analysis,  satisfying the following relations:

\begin{eqnarray*}
	\boldsymbol{\Omega}_{1} & = & \dot{\boldsymbol{v}}\wedge\ddot{\boldsymbol{v}}\\
	& = & \frac{1}{\left(s^{\prime}\right)^{2}}\boldsymbol{v}^{\prime}\wedge\boldsymbol{v}^{\prime\prime}\\
	& = & k_{1}\boldsymbol{e}_{1}\wedge\boldsymbol{e}_{2}\\
	& = & s^{\prime}\left\Vert \ddot{\boldsymbol{v}}\right\Vert \boldsymbol{e}_{1}\wedge\boldsymbol{e}_{2}\\
	& = & \frac{1}{s^{\prime}}\sqrt{\left\Vert \boldsymbol{v}^{\prime\prime}\right\Vert ^{2}-2\frac{s^{\prime\prime}}{s^{\prime}}\left(\boldsymbol{v}^{\prime}\cdot\boldsymbol{v}^{\prime\prime}\right)+\left(s^{\prime\prime}\right)^{2}}\boldsymbol{e}_{1}\wedge\boldsymbol{e}_{2}\\
	\dot{\boldsymbol{v}}\rfloor\left(\boldsymbol{\Omega}-\boldsymbol{\Omega}_{1}\right) & = & \sum_{i=2}^{n-1}\boldsymbol{\Omega}_{i}=0
\end{eqnarray*}

The difference $\boldsymbol{\Omega}-\boldsymbol{\Omega}_{1}$ is proportional
to the bivector $\boldsymbol{B}$ of relation (3.8) in \cite{D.Hestenes1987},
which always satisfies $\dot{\boldsymbol{v}}\rfloor\boldsymbol{B}=0$.
When the curve $\boldsymbol{v}\left(t\right)$ is a planar one (such
as in the case for 3-phase line-to-line voltages signal), all curvature
coefficients are zero except $\kappa_{1}$. In this specific case, we have
$\boldsymbol{\Omega}-\boldsymbol{\Omega}_{1}=0$, and the two quantities
are equivalent $\boldsymbol{\Omega}=\boldsymbol{\Omega}_{1}$. 

\section{Example Signals}
A number of theoretical examples are now presented to validate the proposed method. Sinusoidal and non-sinusoidal multi-phase systems with an arbitrary number of phases are studied. The goal is to obtain a geometric representation of the generalized frequency grid by using the concept of DB.

\subsection{Multi-phase Balanced Sinusoidal Signal}

Assume we have a balanced multi-phase sinusoidal electrical signal. It means that the amplitude is the same for all phases and the phase angle among them is $\frac{2\pi m }{n}$ with $m=0,1,\ldots,n-1$ and $n$ the number of phases. 

\begin{equation}
	\begin{aligned}
			\boldsymbol{v}\left(t\right) & =  V\sum_{m=0}^{n-1}\cos\left(\omega t-2\pi\frac{m}{n}\right)\boldsymbol{\sigma}_{m+1}
		 =  V\sum_{m=0}^{n-1}\left[\cos\left(2\pi\frac{m}{n}\right)\cos\left(\omega t\right)+\sin\left(2\pi\frac{m}{n}\right)\sin\left(\omega t\right)\right]\boldsymbol{\sigma}_{m+1}\\
		& =  \cos\left(\omega t\right)V\left[\sum_{m=0}^{n-1}\cos\left(2\pi\frac{m}{n}\right)\boldsymbol{\sigma}_{m+1}\right]+\sin\left(\omega t\right)V\left[\sum_{m=0}^{n-1}\sin\left(2\pi\frac{m}{n}\right)\boldsymbol{\sigma}_{m+1}\right]\\
		& =  \cos\left(\omega t\right)\boldsymbol{a}+\sin\left(\omega t\right)\boldsymbol{b}\\
	\end{aligned}
\end{equation}

\noindent with

\begin{equation*}
	\begin{aligned}
				\boldsymbol{a} & =  V\sum_{m=0}^{n-1}\cos\left(2\pi\frac{m}{n}\right)\boldsymbol{\sigma}_{m+1}\\
		\boldsymbol{b} & =  V\sum_{m=0}^{n-1}\sin\left(2\pi\frac{m}{n}\right)\boldsymbol{\sigma}_{m+1}
	\end{aligned}
\end{equation*}

In this case the two $n$-dimensional vectors $\boldsymbol{a},\boldsymbol{b}$
are orthogonal with $\left\Vert \boldsymbol{a}\right\Vert ^{2}=\left\Vert \boldsymbol{b}\right\Vert ^{2}=\frac{n}{2}V^{2}$.
This signal describes a perfect circular curve in the plane spanned
by the $n$-dimensional orthogonal vectors $\boldsymbol{a}$, $\boldsymbol{b}$
inside the larger signal space defined by orthonormal basis vectors
$\left\{ \boldsymbol{\sigma}_{i}\right\} _{i=1}^{n}$. We can express
all relevant quantities using vectors $\boldsymbol{a}$ and $\boldsymbol{b}$
as follows:

\begin{eqnarray*}
	\boldsymbol{v}^{\prime} & = & -\omega\left(\sin\left(\omega t\right)\boldsymbol{a}-\cos\left(\omega t\right)\boldsymbol{b}\right)\\
	\boldsymbol{v}^{\prime\prime} & = & -\omega^{2}\left(\cos\left(\omega t\right)\boldsymbol{a}+\sin\left(\omega t\right)\boldsymbol{b}\right)\\
	\left\Vert \boldsymbol{v}^{\prime}\right\Vert ^{2} & = & \frac{\omega^{2}}{2}nV^{2}\\
	s^{\prime}=\left\Vert \boldsymbol{v}^{\prime}\right\Vert  & = & \frac{\omega}{\sqrt{2}}\sqrt{n}V\\
	\boldsymbol{u}_{1}=\dot{\boldsymbol{v}} & = & -\frac{\sqrt{2}}{\sqrt{n}V}\left[\sin\left(\omega t\right)\boldsymbol{a}-\cos\left(\omega t\right)\boldsymbol{b}\right]\\
	\boldsymbol{u}_{2}=\ddot{\boldsymbol{v}} & = & -\frac{2}{nV^{2}}\left[\cos\left(\omega t\right)\boldsymbol{a}+\sin\left(\omega t\right)\boldsymbol{b}\right]\\
	\boldsymbol{e}_{1}=\frac{\boldsymbol{u}_{1}}{\left\Vert \boldsymbol{u}_{1}\right\Vert } & = & -\frac{\sqrt{2}}{\sqrt{n}V}\left[\sin\left(\omega t\right)\boldsymbol{a}-\cos\left(\omega t\right)\boldsymbol{b}\right]\\
	\boldsymbol{e}_{2}=\frac{\boldsymbol{u}_{2}}{\left\Vert \boldsymbol{u}_{2}\right\Vert } & = & -\frac{\sqrt{2}}{\sqrt{n}V}\left[\cos\left(\omega t\right)\boldsymbol{a}+\sin\left(\omega t\right)\boldsymbol{b}\right]\\
	k_{1} & = & s^{\prime}\frac{\left\Vert \boldsymbol{u}_{2}\right\Vert }{\left\Vert \boldsymbol{u}_{1}\right\Vert }=\omega\\
	\boldsymbol{\Omega}_{1} & = & k_{1}\boldsymbol{e}_{1}\wedge\boldsymbol{e}_{2}
	 =  \omega\boldsymbol{e}_{1}\wedge\boldsymbol{e}_{2}
	 =  2\omega\left[\frac{1}{nV^{2}}\right]\boldsymbol{a}\wedge\boldsymbol{b}
\end{eqnarray*}

Note the constant nature of $\left\Vert \boldsymbol{\Omega}_{1}\right\Vert =\omega$
for this signal, which confirms the curve being a perfect circle.

\subsection{Multi-phase Unbalanced Sinusoidal Signal}

Assume we have a general (possibly unbalanced) multi-phase sinusoidal electrical signal. This now means that  the amplitude can be different among phases and phase angle is not regularly spaced by $\frac{2\pi m}{n}$. In this case the voltage vector is:

\begin{equation*}
	\begin{aligned}
	\boldsymbol{v}\left(t\right) & =  \sum_{m=1}^{n}V_{m}\cos\left(\omega t-\varphi_{m}\right)\boldsymbol{\sigma}_{m}=
	 \sum_{m=1}^{n}\left[V_{m}\cos\left(\varphi_{m}\right)\cos\left(\omega t\right)+V_{m}\sin\left(\varphi_{m}\right)\sin\left(\omega t\right)\right]\boldsymbol{\sigma}_{m}\\
	& =  \cos\left(\omega t\right)\left[\sum_{m=1}^{n}V_{m}\cos\left(\varphi_{m}\right)\boldsymbol{\sigma}_{m}\right]+\sin\left(\omega t\right)\left[\sum_{m=1}^{n}V_{m}\sin\left(\varphi_{m}\right)\boldsymbol{\sigma}_{m}\right]\\
	& =  \cos\left(\omega t\right)\boldsymbol{a}+\sin\left(\omega t\right)\boldsymbol{b}\\
	\end{aligned}
\end{equation*}

\noindent with

\begin{equation*}
	\begin{aligned}
	\boldsymbol{a} & =  \sum_{m=1}^{n}V_{m}\cos\left(\varphi_{m}\right)\boldsymbol{\sigma}_{m}\\
\boldsymbol{b} & =  \sum_{m=1}^{n}V_{m}\sin\left(\varphi_{m}\right)\boldsymbol{\sigma}_{m}
	\end{aligned}
\end{equation*}

This signal describes an ellipse in the plane spanned by the $n$-dimensional
vectors $\boldsymbol{a}$, $\boldsymbol{b}$ inside the signal space
defined by orthonormal basis vectors $\left\{ \boldsymbol{\sigma}_{i}\right\} _{i=1}^{n}$.
We can express all relevant quantities using vectors $\boldsymbol{a}$
and $\boldsymbol{b}$ as follows:

\begin{eqnarray*}
	\boldsymbol{v}^{\prime} & = & -\omega\left(\sin\left(\omega t\right)\boldsymbol{a}-\cos\left(\omega t\right)\boldsymbol{b}\right)\\
	\boldsymbol{v}^{\prime\prime} & = & -\omega^{2}\left(\cos\left(\omega t\right)\boldsymbol{a}+\sin\left(\omega t\right)\boldsymbol{b}\right)\\
	\left\Vert \boldsymbol{v}^{\prime}\right\Vert ^{2} & = & \frac{\omega^{2}}{2}g^{2}\\
	s^{\prime}=\left\Vert \boldsymbol{v}^{\prime}\right\Vert  & = & \frac{\omega}{\sqrt{2}}g\\
	g & = & \sqrt{\left(\boldsymbol{b}^{2}-\boldsymbol{a}^{2}\right)\cos\left(2\omega t\right)-2\left(\boldsymbol{a}\cdot\boldsymbol{b}\right)\sin\left(2\omega t\right)+\left(\boldsymbol{b}^{2}+\boldsymbol{a}^{2}\right)}\\
	\boldsymbol{u}_{1}=\dot{\boldsymbol{v}} & = & -\frac{\sqrt{2}}{g}\left[\sin\left(\omega t\right)\boldsymbol{a}-\cos\left(\omega t\right)\boldsymbol{b}\right]\\
	\boldsymbol{u}_{2}=\ddot{\boldsymbol{v}} & = & -\frac{4}{g^{4}}\left[\left(\boldsymbol{b}^{2}\cos\left(\omega t\right)-\left(\boldsymbol{a}\cdot\boldsymbol{b}\right)\sin\left(\omega t\right)\right)\boldsymbol{a}+\left(\boldsymbol{a}^{2}\sin\left(\omega t\right)-\left(\boldsymbol{a}\cdot\boldsymbol{b}\right)\cos\left(\omega t\right)\right)\boldsymbol{b}\right]\\
	\boldsymbol{e}_{1}=\frac{\boldsymbol{u}_{1}}{\left\Vert \boldsymbol{u}_{1}\right\Vert } & = & -\frac{\sqrt{2}}{g}\left[\sin\left(\omega t\right)\boldsymbol{a}-\cos\left(\omega t\right)\boldsymbol{b}\right]\\
	\boldsymbol{e}_{2}=\frac{\boldsymbol{u}_{2}}{\left\Vert \boldsymbol{u}_{2}\right\Vert } & = & -\frac{\sqrt{2}}{g}\left[\frac{\boldsymbol{b}^{2}\cos\left(\omega t\right)-\left(\boldsymbol{a}\cdot\boldsymbol{b}\right)\sin\left(\omega t\right)}{\sqrt{\boldsymbol{a}^{2}\boldsymbol{b}^{2}-\left(\boldsymbol{a}\cdot\boldsymbol{b}\right)^{2}}}\boldsymbol{a}+\frac{\boldsymbol{a}^{2}\sin\left(\omega t\right)-\left(\boldsymbol{a}\cdot\boldsymbol{b}\right)\cos\left(\omega t\right)}{\sqrt{\boldsymbol{a}^{2}\boldsymbol{b}^{2}-\left(\boldsymbol{a}\cdot\boldsymbol{b}\right)^{2}}}\boldsymbol{b}\right]\\
\end{eqnarray*}

\begin{eqnarray*}
	k_{1} & = & s^{\prime}\frac{\left\Vert \boldsymbol{u}_{2}\right\Vert}{\|\bm{u}_1\|} 
 =  \left(\frac{2}{g^{2}}\sqrt{\boldsymbol{a}^{2}\boldsymbol{b}^{2}-\left(\boldsymbol{a}\cdot\boldsymbol{b}\right)^{2}}\right)\omega\\
\boldsymbol{\Omega}_{1} & = & k_{1}\boldsymbol{e}_{1}\wedge\boldsymbol{e}_{2}
 =  \left(\frac{2}{g^{2}}\sqrt{\boldsymbol{a}^{2}\boldsymbol{b}^{2}-\left(\boldsymbol{a}\cdot\boldsymbol{b}\right)^{2}}\right)\omega\boldsymbol{e}_{1}\wedge\boldsymbol{e}_{2}\\
& = & 2\omega\left[\frac{1}{g^{2}}-\left(\boldsymbol{a}\cdot\boldsymbol{b}\right)\frac{\sin\left(2\omega t\right)}{g^{4}}\right]\boldsymbol{a}\wedge\boldsymbol{b}
\end{eqnarray*}
In this signal we have $\left\Vert \boldsymbol{\Omega}_{1}\right\Vert =h\left(t\right)\omega$,
where the constant angular frequency of the grid $\omega$ is scaled
by the periodic time dependent factor: 

\begin{eqnarray*}
	h\left(t\right) & = & \frac{2\sqrt{\boldsymbol{a}^{2}\boldsymbol{b}^{2}-\left(\boldsymbol{a}\cdot\boldsymbol{b}\right)^{2}}}{\left(\boldsymbol{b}^{2}-\boldsymbol{a}^{2}\right)\cos\left(2\omega t\right)-2\left(\boldsymbol{a}\cdot\boldsymbol{b}\right)\sin\left(2\omega t\right)+\left(\boldsymbol{b}^{2}+\boldsymbol{a}^{2}\right)}
\end{eqnarray*}

Which has unit average value $\overline{h}=\frac{1}{T}\int_{0}^{T}h\left(t\right)dt=1$
where $T=\frac{\pi}{\omega}$ is the time of a single cycle of $h\left(t\right)$,
which is half the time of a single cycle of the signal $\boldsymbol{v}\left(t\right)$.
This clearly indicates that by averaging $\boldsymbol{\Omega}=\boldsymbol{\Omega}_{1}$
on one half cycle of this signal, followed by taking the norm of the
resulting bivector, we again get the grid frequency $\omega$ as intuitively
expected.

\subsection{Multi-phase Balanced Harmonic Signal}

Finally, assume we have the following harmonic electrical signal: 

\begin{eqnarray*}
	\boldsymbol{v}\left(t\right) & = & \sqrt{2}\left[200\sin\left(\omega t\right)+20\sin\left(2\omega t\right)-30\sin\left(7\omega t\right)\right]\boldsymbol{\sigma}_{1}\\
	& + & \sqrt{2}\left[200\sin\left(\omega t-\frac{2\pi}{3}\right)+20\sin\left(2\left(\omega t-\frac{2\pi}{3}\right)\right)-30\sin\left(7\left(\omega t-\frac{2\pi}{3}\right)\right)\right]\boldsymbol{\sigma}_{2}\\
	& + & \sqrt{2}\left[200\sin\left(\omega t+\frac{2\pi}{3}\right)+20\sin\left(2\left(\omega t+\frac{2\pi}{3}\right)\right)-30\sin\left(7\left(\omega t+\frac{2\pi}{3}\right)\right)\right]\boldsymbol{\sigma}_{3}
\end{eqnarray*}

This electrical signal traces a planar symmetric curve in the plane
orthogonal to vector $\left(1,1,1\right)$. The expression for
the first DB is:

\begin{eqnarray*}
	\boldsymbol{\Omega}_{1}\left(t\right) & = & \frac{5\omega}{\sqrt{3}}\left(\frac{16\cos\left(3\omega t\right)+672\cos\left(6\omega t\right)+84\cos\left(9\omega t\right)-691}{-160\cos\left(3\omega t\right)+840\cos\left(6\omega t\right)+168\cos\left(9\omega t\right)-857}\right)\left(\boldsymbol{\sigma}_{1,2}-\boldsymbol{\sigma}_{1,3}+\boldsymbol{\sigma}_{2,3}\right)
\end{eqnarray*}

The average DB and its norm are given by:

\begin{eqnarray*}
	\overline{\boldsymbol{\Omega}}_{1} & = & \frac{1}{T}\int_{0}^{T}\boldsymbol{\Omega}_{1}\left(t\right)dt
	 =  \sqrt{3}\omega\left(\boldsymbol{\sigma}_{1,2}-\boldsymbol{\sigma}_{1,3}+\boldsymbol{\sigma}_{2,3}\right)  \qquad T=\frac{2\pi}{\omega}\\\\
	\left\Vert \overline{\boldsymbol{\Omega}}_{1}\right\Vert  & = & 3\omega
\end{eqnarray*}

The norm of the average angular velocity blade $\left\Vert \overline{\boldsymbol{\Omega}}_{1}\right\Vert $
is proportional to the grid nominal  angular  frequency $\omega$.

\section{Conclusion}

This paper has presented the generalized concept of geometrical angular frequency applied to multi-phase systems with arbitrary number of phases, extending previous works where linear algebra were used to define the concept of geometric frequency. Geometric Algebra and Differential Geometry have been used to represent vectors in $n$-dimensional spaces and to compute the geometric invariants associated to the generated spatial curves. Voltage signals have been used to create such vectors and to compute the Darboux Bivector, which encodes the specific differential geometric properties in the curves known as electrical curves. The method can be conveniently employed for a variety of engineering problems such as voltage stability, frequency control, to mention a few.

 \bibliographystyle{splncs04}
\bibliography{ICACGA-paper}

\begin{thebibliography}{10}
\providecommand{\url}[1]{\texttt{#1}}
\providecommand{\urlprefix}{URL }
\providecommand{\doi}[1]{https://doi.org/#1}

\bibitem{Beilina2017}
Beilina, L., Karchevskii, E., Karchevskii, M.: Numerical Linear Algebra: Theory
  and Applications. Springer International Publishing (2017).
  \doi{10.1007/978-3-319-57304-5}

\bibitem{Bjoerck1994}
Björck, {\AA}.: Numerics of gram-schmidt orthogonalization. Linear Algebra and
  its Applications  \textbf{197-198},  297--316 (jan 1994).
  \doi{10.1016/0024-3795(94)90493-6}

\bibitem{buzzi2010interpolation}
Buzzi-Ferraris, G., Manenti, F.: Interpolation and regression models for the
  chemical engineer: Solving numerical problems. John Wiley \& Sons (2010)

\bibitem{Gluck1966}
Gluck, H.: Higher curvatures of curves in euclidean space. The American
  Mathematical Monthly  \textbf{73}(7),  699--704 (aug 1966).
  \doi{10.1080/00029890.1966.11970818}

\bibitem{D.Hestenes1987}
Hestenes, D., Sobczyk, G.: Clifford Algebra to Geometric Calculus. Springer
  Netherlands (Aug 1987)

\bibitem{hitzer2003gram}
Hitzer, E.M.: Gram-schmidt orthogonalization in geometric algebra  (2003)

\bibitem{Hubbard1999}
Hubbard, J.H., Hubbard, B.B.: Vector calculus, linear algebra, and differential
  forms: a unified approach (5th edition). Matrix Editions, Upper Saddle River,
  N.J (2015)

\bibitem{RichardKhoury2016}
Khoury, R., Harder, D.W.: Numerical Methods and Modelling for Engineering.
  Springer International Publishing (May 2016)

\bibitem{kirkham2018defining}
Kirkham, H., Dickerson, W., Phadke, A.: Defining power system frequency. In:
  2018 IEEE Power \& Energy Society General Meeting (PESGM). pp.~1--5. IEEE
  (2018)

\bibitem{leon1995time}
Leon, C.: Time-frequency analysis: theory and applications. USA: Pnentice Hall
  (1995)

\bibitem{milano2021geometrical}
Milano, F.: A geometrical interpretation of frequency. IEEE Transactions on
  Power Systems  \textbf{37}(1),  816--819 (2021)

\bibitem{milano2021applications}
Milano, F., Tzounas, G., Dassios, I., Kerci, T.: Applications of the frenet
  frame to electric circuits. IEEE Transactions on Circuits and Systems I:
  Regular Papers  (2021)

\bibitem{montoya2022geometric}
Montoya, F.G., Ba{\~n}os, R., Alcayde, A., Arrabal-Campos, F.M.,
  Rold{\'a}n-P{\'e}rez, J.: Geometric algebra applied to multiphase electrical
  circuits in mixed time--frequency domain by means of hypercomplex hilbert
  transform. Mathematics  \textbf{10}(9), ~1419 (2022)

\bibitem{montoya2021determination}
Montoya, F.G., De~Leon, F., Arrabal-Campos, F.M., Alcayde, A.: Determination of
  instantaneous powers from a novel time-domain parameter identification method
  of non-linear single-phase circuits. IEEE Transactions on Power Delivery
  (2021)

\end{thebibliography}

\end{document}